\pdfoutput=1

\documentclass[10pt,journal,compsoc]{IEEEtran}

\ifCLASSOPTIONcompsoc
\usepackage[nocompress]{cite}
\else
\usepackage{cite}
\fi

\def\BibTeX{{\rm B\kern-.05em{\sc i\kern-.025em b}\kern-.08emT\kern-.1667em\lower.7ex\hbox{E}\kern-.125emX}}

\usepackage[ruled,vlined,lined,commentsnumbered]{algorithm2e}
\usepackage{balance}
\usepackage{csvsimple}
\usepackage{longtable}
\usepackage{array, tabularx}
\usepackage{booktabs}
\usepackage[skip=1pt,font=small,labelfont=bf,textfont=bf]{caption}
\usepackage{url}
\usepackage{lscape}
\usepackage{rotating}
\usepackage{tikz}
\usepackage[normalem]{ulem}
\usepackage{enumitem}
 \usepackage{array,multirow,graphicx}
 \usepackage{float}
\usepackage{amsmath}
\usepackage{amssymb}
\usepackage{amsthm}
\usepackage{fancybox}
\usepackage{caption}
\usepackage{subcaption}
\usepackage{hyperref}

\usepackage[most]{tcolorbox}

\usepackage{threeparttable}

\usepackage{marvosym}
\usepackage{textcomp}

\usepackage{tikz}

\setlist{noitemsep} 

\def\BibTeX{{\rm B\kern-.05em{\sc i\kern-.025em b}\kern-.08em
    T\kern-.1667em\lower.7ex\hbox{E}\kern-.125emX}}

\newcommand{\find}[1]{
\begin{tcolorbox}[tile,size=fbox,boxsep=2mm,boxrule=0pt,top=-2pt,bottom=-2pt,
borderline west={1.5mm}{0pt}{blue!50!white},colback=blue!5!white]
\small #1
\end{tcolorbox}

}

\definecolor{codegreen}{rgb}{0,0.6,0}
\definecolor{codegray}{rgb}{0.5,0.5,0.5}
\definecolor{codepurple}{rgb}{0.58,0,0.82}
\definecolor{backcolour}{rgb}{0.95,0.95,0.92}
 
\definecolor{DarkOrange}{rgb}{0.8,0.3,0.0} 
\definecolor{yellow}{RGB}{255,255,153}
\definecolor{grey}{RGB}{224,224,224}

\newboolean{showcomments}
\setboolean{showcomments}{false}
\ifthenelse{\boolean{showcomments}}
 { \newcommand{\mynote}[2]{
      \fbox{\bfseries\sffamily\scriptsize#1}
        {\small$\blacktriangleright$\textsf{\emph{#2}}$\blacktriangleleft$}}}
        { \newcommand{\mynote}[2]{}}

\begin{document}

\title{Learning to Catch Security Patches
}

\author{
    \IEEEauthorblockN{Arthur D. Sawadogo\IEEEauthorrefmark{1}, Tegawend\'e F. Bissyand\'e\IEEEauthorrefmark{2}, Naouel Moha\IEEEauthorrefmark{1},  \\Kevin Allix\IEEEauthorrefmark{2}, Jacques Klein\IEEEauthorrefmark{2}, Li Li\IEEEauthorrefmark{3}, Yves Le Traon\IEEEauthorrefmark{2} }\\
    \IEEEauthorblockA{\IEEEauthorrefmark{1}Université du Qu\'ebec \`a Montr\'eal
    \\}
    \IEEEauthorblockA{\IEEEauthorrefmark{2}University of Luxembourg
    \\}
    \IEEEauthorblockA{\IEEEauthorrefmark{3}Monash University
    \\}
}

\IEEEtitleabstractindextext{%
\begin{abstract}
Timely patching is paramount to safeguard users and maintainers against dire consequences of malicious attacks. In practice, patching is prioritized following the nature of the code change that is committed in the code repository. When such a change is labeled as being security-relevant, i.e., as fixing a vulnerability, maintainers rapidly spread the change and users are notified about the need to update to a new version of the library or of the application. Unfortunately, oftentimes, some security-relevant changes go unnoticed as they represent {\em silent fixes} of vulnerabilities. In this paper, we propose a Co-Training-based approach to catch security patches as part of an automatic monitoring service of code repositories. Leveraging different classes of features, we empirically show that such an automation is feasible and can yield a precision of over 90\% in identifying security patches, with an unprecedented recall of over 80\%. Beyond such a benchmarking with ground truth data which demonstrates an improvement over the state-of-the-art, we confirmed that our approach can help catch security patches that were not reported as such.
\end{abstract}


}
\maketitle

\section{Introduction}
\label{sec:introduction}


\IEEEPARstart{I}n the last couple of years, our digital world was shaken by two of the most widespread malware outbreaks to date, namely WannaCry and Petya.
Interestingly, both leveraged a known exploit with an available patch~\cite{trendPatching}. Despite the availability of such a patch that could have prevented an infection, a large number of systems around the globe were impacted, leading to a loss of over 4 billion US dollars~\cite{cbsWannacry}. In a typical scenario of vulnerability correction, a developer proposes changes bundled as a software {\em patch} by pushing a {\em commit} (i.e., patch + description of changes) which is analyzed by the project maintainer, or a chain of maintainers, who eventually reject or apply the changes to the master branch. When the patch is accepted and released, all users of the relevant code must apply it to limit their exposure to attacks. The reality, however, is that, for most organizations, there is a lag between a patch release and its application. While in the cases of critical systems, maintainers are hesitant to deploy updates that will hinder operations with downtime, in many other cases, the lag is due to the fact that the proposed change has not been properly advertised as {\em security-relevant}, and is not thus viewed as critical.

Patching is an absolute necessity. Timely patching of vulnerabilities in software, however, mainly depends on the tags associated to the change, such as the commit log message, or on the availability of references in public vulnerability databases. For example, nowadays, developers and system maintainers rely on information from the National Vulnerability Database~\cite{nist} to react to all disclosed vulnerabilities.  Unfortunately, a recent study on the state of open source security~\cite{snyk} revealed that only 9\% of maintainers file for a Common Vulnerability Enumeration (CVE) ID after releasing a fix to a vulnerability. The study further reports that 25\% of open source software projects completely silently fix vulnerabilities without disclosing them to any official repository.

Silent vulnerability fixes are a concern for third-party developers and users alike. Given the low coverage of official vulnerability repositories, there are initiatives in the software industry to automatically and systematically monitor source code repositories in real-time for identifying security-relevant commits, for example by parsing the commit logs~\cite{zhou2017automated}. Manual analysis of code changes is indeed heavy in terms of manpower constraints, requires expert knowledge, and can be error-prone.

\find{Our work deals with the automation of the identification of security patches (i.e., patches fixing vulnerabilities) once a commit is contributed to a code base. To align with realistic constraints\footnote{In practice, identifying security patches must be done at commit-time. An approach would be very successful if it could leverage future comments of bug reports and advisories inputs (e.g., CVE). Such information is however not available in reality when the commit is made.} of practitioners, we only leverage information available within the commit.}

In this paper, we investigate the possibility to apply machine learning techniques to automate the identification of source code changes that actually represent security patches. To that end, we investigate three different classes of features related to the change metadata (e.g., commit logs), the code change details (e.g., number of lines modified), as well as specific traits that are recurrent in vulnerabilities (e.g., array index change). We then build on the insight that analysts can {\em independently} rely either on commit logs or on code change details to suspect a patch of addressing a vulnerability. Thus, we propose to build a Co-Training approach where two classifiers leverage separately text features and code features to eventually learn an effective model. This semi-supervised learning approach further accounts for the reality that the datasets available in practice include a {\em large portion of samples whose labels (i.e., ``security-relevant'' or not) are unknown}.
Overall, we make the following contributions:
\begin{itemize}[leftmargin=*,noitemsep,topsep=0pt]
	\item We motivate and dissect the problem of identifying security-relevant code changes. In particular, we investigate the discriminative power of a variety of features to clarify the possibility of a learning process.
	\item We propose a semi-supervised approach with Co-Training~\cite{blum1998combining} which we demonstrate to yield high precision (95\%) and recall (88\%). This represents a significant improvement over the state-of-the-art.
	\item Finally, we show that our approach can help flag patches that were unlabeled until now. We have confirmed our findings by manual analysis, with the help of external expertise. 
\end{itemize}

The remainder of this paper is organized as follows. We motivate our study and
enumerate related work in Section~\ref{sec:related}.  Section~\ref{sec:approach} describes our approach while Section~\ref{sec:assessment} presents the experimental study and results. Section~\ref{sec:discussion} discusses threats to validity and future work.  Section~\ref{sec:conclusion} concludes this work.

\section{Motivation \& Related Work}
\label{sec:related}
The identification of security-relevant commits has applications for various stakeholders in software development.
The literature includes a number of related work that we summarize in this section.

\subsection{Motivating cases}
The urgency of updating a software given a proposed change is assessed at different levels of the software development cycle. We consider the cases of developer-maintainer and maintainer-user communications.

\noindent
{\bf (1) \em Patch processing delays by maintainers.}
We consider the case of the Linux kernel, which is developed according to a hierarchical open
source model referred to as Benevolent dictator for life (BDFL)~\cite{van2008origin}. In this model,
 anyone can contribute, but ultimately all contributions
are integrated by a single person, Linus Torvalds, into the mainline development tree. A Linux kernel
maintainer receives patches related to a particular file or subsystem
from developers or more specialized maintainers. After evaluating
and locally committing them, he/she propagates them upwards in
the maintainer hierarchy, eventually up to Linus Torvalds. 
Since the number of maintainers is significantly lower than that of contributors, there is a delay between a patch authoring date and its commit date. A recent study, however, has shown that author patches for Linux are addressed in a timely manner by maintainers~\cite{koyuncu2017impact}. Nevertheless, given the critical nature of a security patch, we expect its processing to be even more speedy if the commit message contains relevant information that attracts maintainers' attention.

\begin{figure}[!h]
\centering
\includegraphics[width=0.75\linewidth]{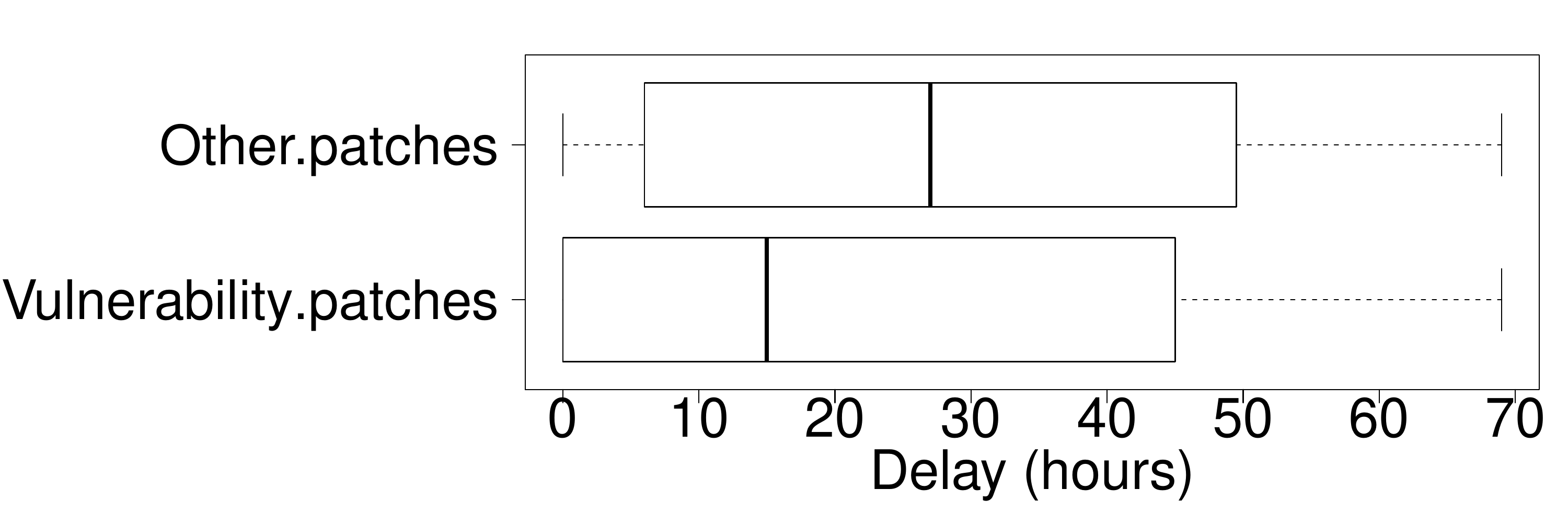}
\caption{Delays for validating contributor patches in Linux}
\label{fig:commit-delays}
\vspace{-0.5cm}
\end{figure}

Figure~\ref{fig:commit-delays} illustrates the delay computed on randomly sampled sets of 1\,000 commits where the log clearly contained a CVE reference, and 1\,000 commits with no such references. The delay is computed as the difference of time between the contribution date (i.e., Author date in git) and the date it was accepted in the repository (i.e., Commit date in git). The boxplots show how patches that are explicitly related to vulnerabilities are validated faster than other patches: on median average, security patches are validated fifteen hours faster. We confirmed that the difference is statistically significant with MWW tests~\cite{mann1947test}. 

\vspace{-5mm}

\find{If proper notice is given, maintainers are likely to prioritize the validation and propagation of security patches.}

\noindent
{\bf (2) \em Version release delays for users.}
In the development cycle of software, versioning allows maintainers to fix milestones with regards to the addition of new features, or the stabilization of a well-tested branch after the application of several bug fixes. However, when a security patch is applied to the code base, it is common to see maintainers release a new version early to protect users against potential attacks. To confirm that this is indeed common, we consider the case of the OpenSSL library and compare the delay between a given commit and the subsequent version release date (which is inferred by checking commits with version tags). 
The delay was computed for all the 1\,550 OpenSSL commits (495 of which carry 
security patches) collected in our study datasets. 

\begin{figure}[!h]
\centering
\includegraphics[width=0.7\linewidth]{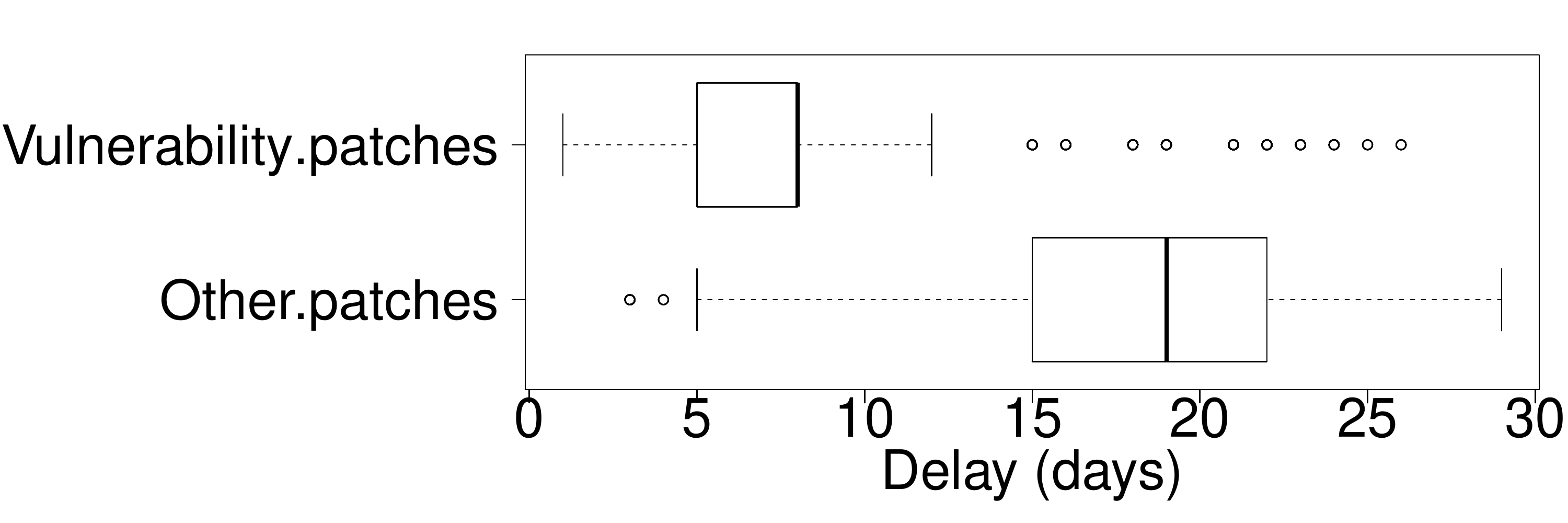}
\caption{Comparative delays for OpenSSL release after a security patch vs after any other patch}
\label{fig:versioning-delays}
\vspace{-4.5mm}
\end{figure}

Boxplot representations in Figure~\ref{fig:versioning-delays} show that many OpenSSL versions are released just after security patches. In contrast, the gap between any other commit and a version release is bigger: releases are made on average seven days after a security patch, but about twenty days after other types of patches.

\vspace{-4.5mm}

\find{To reduce user exposure, it is necessary to release new versions when vulnerabilities are patched. To that end, it is critical to identify such security patches.}
%

\vspace{-1.1mm}
\subsection{Related work}
Our work is related to several research directions in the literature, most notably studies on 1) vulnerability management, 2) the application of machine learning in software maintenance tasks, 3) change analysis, and 4) commit classification.

\vspace{-1mm}
\subsubsection{Vulnerability management} Recently, the topic of Autonomous Cyber Reasoning System~\cite{ji2018coming} has attracted extensive attention from both industry and academia, with the development of new techniques to automate the detection, exploitation, and patching of software vulnerabilities in a scalable and cost-effective way. Static analysis approaches such as the code property graph by Yamaguchi et al.~\cite{yamaguchi2014modeling} require a built model of vulnerabilities based on expert knowledge. Dynamic approaches leverage fuzzing to test a software with intentionally invalid inputs to discover unknown vulnerabilities~\cite{godefroid2008automated,sutton2007fuzzing}, or exploit taint analyses to track marked information flow through a program as it executes in order to detect most types of vulnerabilities~\cite{newsome2005dynamic}, including leaks~\cite{li2015iccta}. Such approaches, although very precise, are known to be expensive, and achieve a limited code coverage~\cite{brooks2017survey}. Recently, researchers have been investigating concolic analysis~\cite{cadar2008klee} tools for software security. Mayhem~\cite{cha2012unleashing} is an example of such a system. 

The literature includes a number of approaches that use software metrics to highlight code regions that are more likely to contain vulnerabilities. Metrics such as code churn and code complexity along with organizational measures (e.g., team size, working hours) allowed to achieve high precision in a large scale empirical study of vulnerabilities in Windows Vista~\cite{zimmermann2010searching}. However, Jay et al.~\cite{jay2009cyclomatic} have warned that many of these metrics may be highly correlated with lines of code, suggesting that such detection techniques are not helpful in reducing the amount of code to read to discover the actual vulnerable piece of code.

Nowadays, researchers are exploring machine learning techniques to improve the performance of automatic software vulnerability detection, exploitation, and patching~\cite{ji2018coming,li2018vuldeepecker}. For
example, Scandariato et al.~\cite{scandariato2014predicting} have trained a classifier on textual
features extracted from source code to determine vulnerable
software components. Several unsupervised learning approaches have been presented to assist
	in the discovery of vulnerabilities~\cite{yamaguchi2013chucky,chang2008discovering}. We differ from these approaches both in terms of objectives, and in the use of a combination of features from code and metadata. With respect to feature learning, new deep learning-based approaches~\cite{li2018vuldeepecker} are being proposed since they do not require expert intervention to generate features. The models are however mostly opaque~\cite{PontinWired18} for analysts who require explainability of decisions during audits. Finally, it is noteworthy that the industry is starting to share with the research community some datasets yielded by manual curation efforts of security experts~\cite{ponta2019manually}.

\vspace{-1mm}
\subsubsection{Machine learning for software maintenance} The research on applying machine learning to automate software maintenance tasks has been very active in recent years. We refer the reader to a recent survey by Allamanis et al.~\cite{allamanis2018survey}. In such approaches, researchers rely on the {\em naturalness}~\cite{hindle2012naturalness} of software code to build prediction models. Our work also falls under these categories as we build on the assumption that security patches have discriminating features with respect to other patches (whether bug fix patches or enhancement patches).

\vspace{-1mm}
\subsubsection{Change analysis}
Software change is a fundamental ingredient of software maintenance~\cite{li2013survey}. Software changes are often applied to comply to new requirements, to fix bugs, 
to address change requests, and so on. When such changes are made, inevitably, some expected and unexpected effects may ensue, even beyond the software code. Software change impact analysis has been studied 
in the literature as a collection of techniques for determining the effects of the proposed changes on other
parts of the software~\cite{arnold1996software}. 

Researchers have further investigated a number of prediction approaches related to software changes, including by analysing co-change patterns to predict source code changes~\cite{ying2004predicting}. Closely related to ours is the work of Tian et al.~\cite{tian2012identifying} who propose a learning model to identify Linux bug fixing patches. The motivation of their work is to improve the propagation of fixes upwards the mainline tree. 
Our approach, however, is substantially different regarding: 
(1) \textit{Objective}: \cite{tian2012identifying} targets Linux development, and identifies bug fixes. We are focused on security patches.
(2) \textit{Method}: \cite{tian2012identifying} leverages the classification algorithm named Learning from Positive and Unlabeled Examples (LPU)~\cite{lpu}. In contrast, we explore Co-Training which requires two independent views of the data. We also include a more security-sensitive set of features. We explore a combination of latent (e.g., \#sizeof) and explicit (e.g., keyword) features.
(3) \textit{Evaluation}: \cite{tian2012identifying} was evaluated against a keyword-based approach. We evaluate against the state-of-the-art and based on manual audit. All data is released and made available for replication. Following up on the work of Tian et al.~\cite{tian2012identifying}, Hoang et al. have proposed a deep learning-based tool for classifying bug fix commits~\cite{Hoang18}. 

Security analysis of commits has been investigated by Perl et al.~\cite{perl2015vccfinder} who presented VCCFinder for flagging suspicious commits by using an SVM classifier. In contrast to our work, VCCFinder aims at identifying vulnerability-introducing changes, while, conversely, we aim for identifying those changes that fix vulnerabilities. 

\vspace{-1mm}
\subsubsection{Commit classification}
Recently, researchers from the security industry~\cite{zhou2017automated, sabetta2018icsme} (from SourceClear, Inc. and SAP respectively)
have presented early investigations on the prediction of security issues in relation with commit changes.
Zhou and Asankhaya~\cite{zhou2017automated} focus on commit logs, commit metadata and associated bug reports, and leverage regular expressions to identify features for predicting security-relevant commits. The authors use embedding ($word2vec$) to learn the features, which leads to an opaque decision-making system~\cite{PontinWired18,KnightMIT17} when it comes to guiding a security analyst in his/her auditing tasks. The approach is further limited since experimental data show that not all fixes are linked to reported bugs, and not all developers know (or want to disclose in logs) that they are fixing vulnerabilities.
Sabetta and Bezzi~\cite{sabetta2018icsme} improve over the work of Zhou and Asankhaya by considering code changes as well. Their approach is fully-supervised (thus, assuming that the labeled dataset is perfect and sufficient).

\find{ Closely related work in identifying security patches are contributed so far by the industry. Nevertheless, various academic works rely on scarce data to train machine learning models for vulnerability detection, exploitation or patching. Our work will enable the amplification of such datasets (beyond the disclosed security patches), to include silent fixes, thus increasing the coverage and reliability of the state-of-the-art. }

\vspace{-0.5mm}
\section{Approach}
\label{sec:approach}
\vspace{-0.5mm}

Our work addresses a {\bf binary classification problem} of distinguishing security patches from other patches: we consider a combination of {\em text analysis of  commit
logs} and {\em code analysis of commit changes diff} to catch security patches.
To that end, we proceed to the extraction of ``facts'' from text and code, and then perform 
a feature engineering that we demonstrate to be efficient for discriminating security patches from other patches. Finally, we learn a prediction model using machine learning classification techniques.

In a typical classification task, an appropriately labeled training
dataset is available. In our setting, however, this is not  the case as introduced earlier: in our dataset, when a commit is attached to a CVE, we can guarantee that it does provide a security patch; when the commit does not mention a CVE, we cannot assume that it does not provide a security patch. Therefore, for positive data, i.e., security patches, we
can leverage the limited dataset of patches that have been listed in vulnerability databases (e.g., the NVD). 
There is, however, no corresponding set of independently labeled
negative data, i.e., non-security patches, given that developers may silently fix their vulnerable code. 
This problem was raised in previous work on the identification of bug fixing patches by Tian et al.~\cite{tian2012identifying}. Nevertheless, our setting requires even more refined analysis since security patches can be easily confused with a mere non-security-relevant bug fix. To address the problem of having a small set of labeled data and a large set of unlabeled data for security patches, we consider a Co-Training~\cite{blum1998combining} approach where 
we combine two models, each trained with features 
extracted from two disjoint aspects (commit message vs. code diff) of our dataset.
This process has been shown to be one of the most effective techniques for semi-supervised learning~\cite{nigam2000analyzing}. 

\vspace{-1.0mm}
\find{Concretely, our Co-Training approach considers commit logs, on the one hand, and code diffs, on the other hand, as redundant views of the changes, given that the former describes the latter. Then we train two separate classifiers, one for each view, that are iterated by exchanging labeled data until they agree on classification decisions (cf. Section~\ref{subsec:Co-Training}).}
\vspace{-2.0mm}

In this section, we first provide information on the data acquisition (cf. Section~\ref{subsec:acquisition}), on feature engineering (cf. Section~\ref{subsec:extraction}) and assessment (cf. Section~\ref{subsec:selection}). Then, we present the Co-Training approach (cf. Section~\ref{subsec:Co-Training}).

\vspace{-2.0mm}
\subsection{Data Collection}
\label{subsec:acquisition}
For most modern software, developers rely on the git version control system.
Git makes available the history of changes that have been made to the code base in the form of a series of patches. Thus, a patch constitutes a thorough summary of a code change, describing the modification that a developer has made to the source code at the time of a commit. Typically, a patch as depicted in Figure~\ref{fig:diff}, includes two artifacts: a) the log message in which the developer describes the change in natural language; b) the diff which represents the changes that are to be applied. The illustrated vulnerability, as in many cases, is due to a missing constraint that leaves a window for attackers to exploit.

\begin{figure}[!htb]
	{\parbox{\linewidth}{
\begin{lstlisting}[linewidth={\linewidth},frame=tb,basicstyle=\scriptsize\ttfamily]
commit 5ebff5337594d690b322078c512eb222d34aaa82
Author: Michal Schmidt <anonymized@redhat.com>
Date:   Fri Mar 2 10:39:10 2012 +0100

    util: never follow symlinks in rm_rf_children()
    The function checks if the entry is a directory 
    before recursing, but there is a window between 
    the check and the open, during which the
    directory could be replaced with a symlink.
    CVE-2012-1174 
    https://bugzilla.redhat.com/show_bug.cgi?id=803358

diff --git a/src/util.c b/src/util.c
index 20cbc2b0d..dfc1dc6b8 100644
--- a/src/util.c
+++ b/src/util.c
@@ -3593,7 +3593,8 @@ static int rm_rf_children(int fd,...) {
if (is_dir) {
    int subdir_fd;
-   if((subdir_fd = openat(fd, de->d_name, O_RDONLY|...)) < 0){
+       subdir_fd = openat(fd, de->d_name, O_RDONLY|...|O_NOFOLLOW);
+       if (subdir_fd < 0) {
            if (ret == 0 && errno != ENOENT)
                ret = -errno;
            continue;

\end{lstlisting}
	}}%
	\caption{Example of a security patch in the OpenSSL library}
	\label{fig:diff}
\end{figure}

For our experiments, we consider three projects whose code is widespread among IT systems: the {\bf Linux} kernel development project, the {\bf OpenSSL} library project and the {\bf Wireshark} network protocol analyzer. For each project, we attempt to collect \underline{\em positive} and \underline{\em negative} data for the classical binary classification task, as well as the \underline{\em unlabeled} data for our semi-supervised learning scenario:
\begin{itemize}[leftmargin=*]
\item {\bf Positive data} (i.e., {\em security patches}). We collect patches reported as part of security advisories, and thus known to be addressing a known and reported vulnerability. 
\item {\bf Negative data} (i.e., {\em non-security patches}). We use heuristics to build the dataset of negative data. To ensure that it is unbiased and representative, we explicitly consider different cases of non-security patches, and transparently collect these sets separately with a clear process to enable replication. Concretely, we consider:
\begin{itemize}
	\item \emph{Pure bug fixing patches}. We collect patches that are known to fix bugs in project code, but that are not security-relevant.
	\item \emph{Code enhancement patches}. We collect patches that are not about fixing bugs or vulnerabilities. Such patches may be delivered by commits to perform code cleaning, feature addition, performance enhancement, etc.
\end{itemize}
	\item {\bf Unlabeled data}. We finally collect patches that are about fixing the code, but for which we do not yet know whether it is about fixing a vulnerability or non-security bugs.
\end{itemize}

\begin{figure}[!h]
\centering
\includegraphics[width=1.00\linewidth]{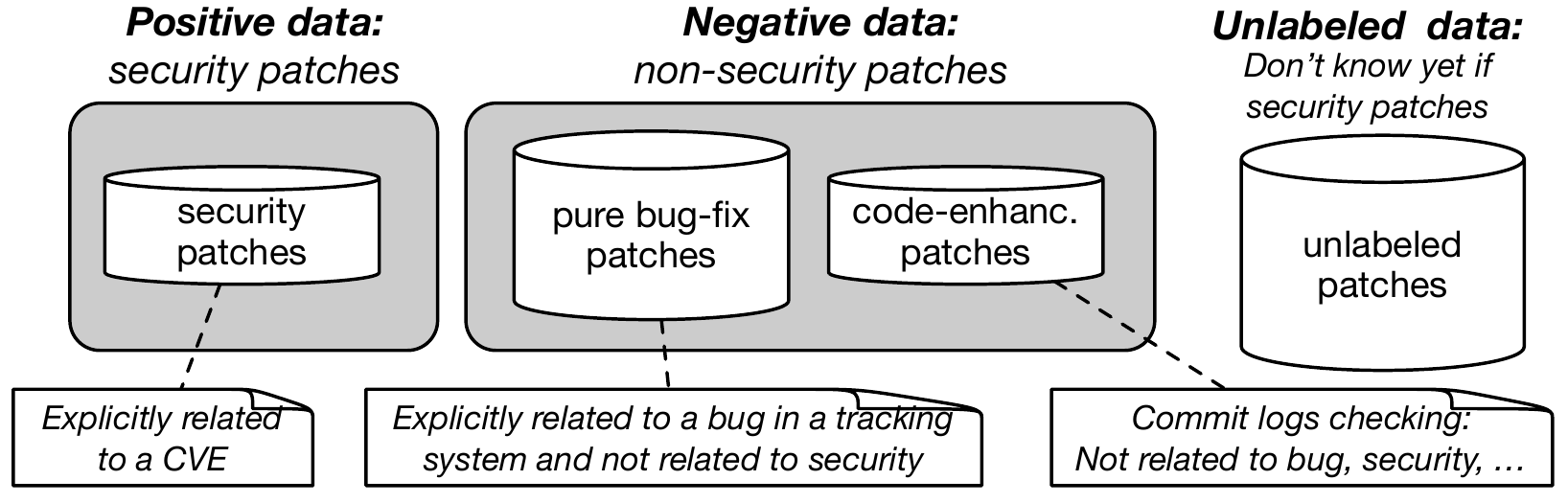}
\caption{Distinct subsets of the dataset built for our experiments}
\label{fig:datasets}
\vspace{-5.0mm}
\end{figure}

The creation of these datasets is summarized in Figure~\ref{fig:datasets} and detailed in the following paragraphs.

\vspace{-4.0mm}

\vspace{-2.0mm}
\subsubsection{Security patches (for positive datasets)}
We acquire security patches by leveraging a recent framework proposed by Jimenez et al.~\cite{jimenez2018enabling} for automated collection of vulnerability instances from software archives. The framework builds upon the National Vulnerability Database information and attempts to connect such information with other sources such as bug tracking systems and git repositories.
The data recovered include information, for each item, about the CVE ID, the CVE description, the time of creation, the associated bug ids from the project bug tracking system, the list of impacted software versions, and the list of commits that fixed the vulnerability.
Overall, as of July 2018, we managed to retrieve 1\;398, 986, and 495 security patches for Linux, Wireshark, and OpenSSL respectively.

\subsubsection{Pure bug fixing patches (for negative datasets)}
To ensure that our approach can effectively differentiate security-relevant fixes from other fixes, we set to collect a dataset of non-security-relevant  patches following conservative heuristics. First, we consider patches that are not reported in a security advisory, and whose commit logs do not include ``vulnerability'' or ``security'' keywords. Then, we focus on those patches whose commits are linked to a bug reported in a bug tracking system. Finally, we ensure that the bug report itself does not hint at a potential security issue. 
For that, we follow the approach proposed by security analysts Zhou and Asankhaya~\cite{zhou2017automated}, 
and drop all cases where the bug report matches the regular expression provided in Table~\ref{tab:regexp}. 
Overall, with this method, we managed to retrieve 1\;934 and 2\;477 pure bug fixing patches for Linux, Wireshark respectively. Our dataset does not contain any pure bug-fix patches for OpenSSL due to missing links between commits and bug reports of OpenSSL. Future work could consider using state-of-the-art bug linking approaches~\cite{nguyen2012multi,wu2011relink,bissyande2013empirical}. 

\begin{table}[H]
\vspace{-2.5mm}
\caption{Regular expression used to filter out security-related issues described in bug reports}
\begin{tabular}{p{7cm}}
\scriptsize
\tt
	(?i)(denial.of.service|\textbackslash bXXE\textbackslash b|remote.code.execution
	|\textbackslash bopen.redirect|OSVDB|\textbackslash vuln|\textbackslash CVE\textbackslash b|\textbackslash bXSS\textbackslash b|\textbackslash bReDoS\textbackslash b
	|\textbackslash bNVD\textbackslash b|malicious|x-frame-options|attack|cross.site 
	|exploit|directory.traversal|\textbackslash bRCE\textbackslash b|\textbackslash bdos\textbackslash b|\textbackslash bXSRF\textbackslash b
	|clickjack|session.fixation|hijack|advisory|insecure |security|\textbackslash bcross-origin\textbackslash b|unauthori[z|s]ed
	|infinite.loop|authenticat(e|ion)|brute\,force|bypass
	|constant.time|crack|credential|\textbackslash bDoS\textbackslash b|expos(e|ing)
	|hack|harden|injection|lockout|overflow|password
	|\textbackslash bPoC\textbackslash b|proof.of.concept|poison|privilege 
	|\textbackslash b(in)?secur(e|ity)|(de)?serializ|spoof|timing|traversal)
	\\
\end{tabular}
\label{tab:regexp}
\vspace{-4.5mm}
\end{table}

\subsubsection{Code enhancement patches (for negative datasets)}
To ensure that our model will not be overfitted to the cases of fixing patches, we collect noise dataset represented by commits that enhance the code base with new feature additions. We thus set to build a parser of commit logs for identifying such commits. To that end, we first manually investigate a small set of 500 commits over all the projects and attempt to identify what keywords can be leveraged. Given the diversity of fixes and commit log tokens, we eventually decide to focus on keywords recurrent in all commits that are not about feature addition, in order to reduce the search space. These are: \emph{bug, fix, bugzilla, resolve, remove, merge, branch, conflict, crash, debug}. Excluding known security patches, known bug fixes (whether pure or not), and those that match the previous keywords, we consider the remaining patches as the sought noise for the learning process. Overall, we collected 681, 658, and 679 code enhancement patches for Linux, Wireshark, and OpenSSL respectively.

\subsubsection{Unlabeled patches}
Ultimately, our goal is to provide researchers and practitioners with an approach for identifying silent security fixing patches. Thus, we hypothesize that some fixing patches are actually unlabeled security patches. To build a dataset of unlabeled patches where security patches may be included, we parse all remaining patches (i.e., patches that are not collected in the previous datasets) and further hone in the subset of unlabeled patches that are more relevant to be caught as security patches. To that end, we focus on commits whose logs match the regular expression  {\tt (?i)(bug|vuln\footnote{Commits with logs matching keyword ``vuln'' cannot be directly considered to be security patches without an audit of the full description and even of the code change.}|fix)}. Eventually, we collected 147\;746, 18\;067, and 437 unlabeled patches for Linux, Wireshark, and OpenSSL respectively.

Table~\ref{tab:dataset} summarizes the statistics on the collected datasets. We note that, as we postulated, most patches are unlabeled. Security patches are mostly silent~\cite{snyk}. Even in the case where a patch is present in a security advisory (i.e., the NIST vulnerability database  in our case), the associated commit log may not explicitly use terms that  hint to a security issue. For example, with respect to the regular expression in Table~\ref{tab:regexp}, we note that 15.21\% of Wireshark security patches, 37.19\% of Linux security patches and up to 98.78\% of OpenSSL security patches do not match security-related tokens.

\begin{table}[!h]
\vspace{-2.0mm}
\centering
\caption{Statistics on the collected datasets}
\resizebox{0.9\linewidth}{!}{
	\begin{tabular}{lrrrr}
	& OpenSSL & Wireshark & Linux & Total\\
	\toprule
	Security patches & 495 &1\;398 & 986 & 2\;879\\
	\midrule
	Pure bug fixing patches &(--) \footnotemark& 1\;934& 2\;477 & 4\;411\\
	\midrule
	Code enhancement patches & 618&681 &658 & 1\;957\\		
	\midrule
	Unlabeled patches &437 & 18\;067&147\;746 & 166\;250\\		
	\bottomrule
	\end{tabular}
	}
	\label{tab:dataset}
\vspace{-5.0mm}
\end{table}
\footnotetext{No pure bug fixing dataset because of links  missing between bugs and commits.}

\subsection{Feature Extraction and Engineering}
\label{subsec:extraction}
The objective of the feature extraction step is to transform the high-volume raw data that we have previously collected into a reduced dataset that includes only the important facts about the samples. The feature extraction then considers both the textual description of the commits (i.e., the message describing the purpose of the change) and the code diff (i.e., the actual modifications performed). The feature engineering step then deals with the representation of the extracted facts into numerical vectors to be fed to machine learning algorithms.

\subsubsection{Commit text features} We extract text features by considering all commit logs as a bag of words, excluding stop words (e.g., ``as'', ``is'', ``would'', etc.) which are very frequently appearing in any English document and will not hold any discriminative power. We then reduce each word to its root form using Porter' stemming~\cite{porter1980algorithm} algorithm. Finally, given the large number of rooted words, and to limit the curse of dimensionality, we focus on the top 10 of the most recurring words in commit logs of security patches for the feature engineering step. This number is selected as a reasonable vector size to avoid having a too-sparse vector for each commit, given that commit logs are generally short. We calculate the {\em inverse document frequency} ({\em idf}), whose formula is provided in the equation below. It is a measure of how much information the word provides, that is, whether it is common or rare across all commit logs. The feature value for each commit is then computed as the $
	idf_{i} =log \frac{|D|}{|\{d_{j} :t_{i} \in d_{j}\}|} $
	with 
$|D| $ being the total number of documents in the corpus and $|\{d_{j} :t_{i} \in d_{j}\}| $ being the number of documents where  term t$_i$ appears.

\subsubsection{Commit code features} Besides description logs, code change details are available in a commit and can contribute to improve the efficiency of the model as demonstrated by Sabetta and Bezzi~\cite{sabetta2018icsme}. Nevertheless in their work, these security researchers considered all code change tokens as a bag of tokens for embedding. In our work, we propose to refine the feature selection by selecting meaningful facts from code to produce an {\em accurate} and {\em explainable} model. To that end, on the one hand, we are inspired by the classification study of Tian et al.~\cite{tian2012identifying}, and we extract code facts representing the spread of the patch (e.g., the number of files/lines modified, etc.), the code units involved (e.g., the number of expressions, boolean operators, function calls, etc.). On the other hand, we manually investigated a sample set of 300 security patches and noticed a few recurring code facts: for example, {\tt sizeof} is often called to fix buffer overflow vulnerabilities, while {\tt goto}, {\tt continue} or {\tt break} constructs are frequently involved in security fixes related to loops, etc. Thus, we engineer two sub-categories of features: {\em code-fix features} and {\em security-sensitive features}. 

Overall, Table~\ref{tab:features} provides an enumeration of the exhaustive list of features used in this study.

\begin{table}[!h]
\vspace{-4.0mm}
\caption{Exhaustive list of features considered for learning}
\centering
\resizebox{1.0\linewidth}{!}{%
\begin{tabular}{ll|ll}
\toprule
ID    &    code-fix features & ID    &    security-sensitive features \\
\toprule
F1 &\#files changed in a commit & F1       &     \#Sizeof added \\
F2  &  \#Loops added & F2       &      \#Sizeof removed  \\
F3 &  \#Loops removed & F3       &       F1 - F2\\
F4  & F2 - F3 & F4        &      F1 + F2\\ 
F5  & F2 + F3 & F5-F6     &       Similar to F1 to F2 for \#continue\\
F6-F9        &      Similar to F2 to F5 for \#ifs & F7-F8     &       Similar to F1 to F2 for \#break\\
F10-F13     &     Similar to F2 to F5 for \#Lines & F9-F10     &       Similar to F1 to F2 for \#INTMAX\\
F14-F17    &      Similar to F2 to F5  & F11-F12    &    Similar to F1 to F2 for \#goto\\
 & for \#Parenthesized expressions & & \\
F18-F21     &     Similar to F2 to F5  & F13-F14    &    Similar to F1 to F2 for \#define\\
 & for \#Boolean operators & & \\
F22-F25     &     Similar to F2 to F5  & F15-F18    &    Similar to F1 to F4 for \#struct\\
& for \#Assignments & & \\
F26-F29    &      Similar to F2 to F5  & F19-F20    &    Similar to F1 to F2 for \#offset\\
 & for \#Functions call & & \\
F30-F33      &    Similar to F2 to F5 for \#Expression &F21-F24    &    Similar to F1 to F4 for \#void\\

\toprule
\toprule
ID    &    text features\\
\toprule
W1-W10      &  10 Most recurrent non-stop words \\
\bottomrule
\end{tabular}
}
\label{tab:features}
\vspace{-5.0mm}
\end{table}

 \vspace{-2.0mm}
\subsection{Feature Assessment}
\label{subsec:selection}

\subsubsection{Statistical analysis}

\begin{figure*}[!t]
\centering
	\includegraphics[width=0.8\linewidth]{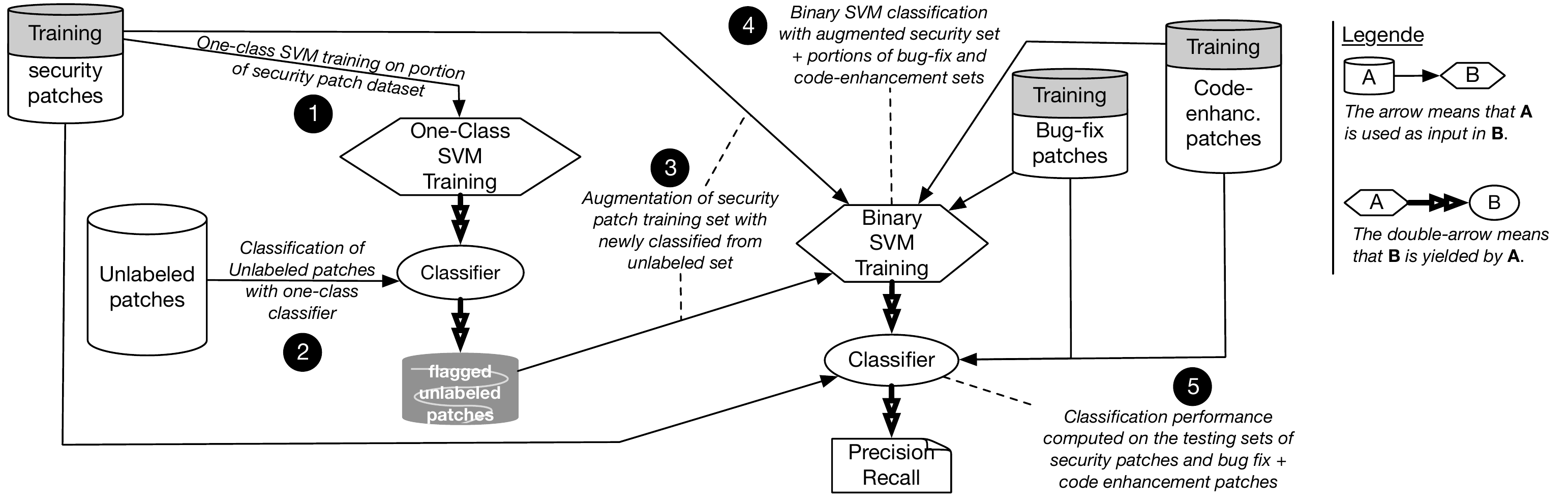}
	\caption{Workflow for assessing the discriminative power of features}
	\label{fig:workflow}
	\vspace{-3.0mm}
\end{figure*}

Before leveraging the features that we have engineered based on manual analysis and intuitive facts,
we propose to assess their fitness with respect to discriminating security patches against
other types of patches. To that end, we used the Mann-Whitney U test~\cite{mann1947test} 
in order to compare the distribution of a given feature within the set
of security patches against the combined set of pure bug fixing patches and code enhancement patches.
The null hypothesis states that the feature
is distributed independently from whether the commit fixes a vulnerability or not. If we can reject the null hypothesis,
the feature is distributed differently in each set and thus
is a promising candidate as input for the machine learning
algorithms.

The Mann-Whitney U tests helped discover that a large majority (i.e., 53 out of 67) of the computed features were not meaningful unless we rescaled the feature values according to the size of the patches. Indeed, for example, code enhancement patches which can be huge (e.g., addition of a new program file) may include a number of loops and sizeof calls, making related features meaningless, unless their numbers are normalized to the size of code in the patch. We then applied, for each feature value per patch, the following formula: 
\vspace{-2.0mm}
\begin{equation}
\small
	F_{norm}= \frac{F}{\# patch\_added\_lines + \#patch\_removed\_lines}
\end{equation}
where the normalized value $F_{norm}$ of a feature is computed by taking into account the patch size.
 Table~\ref{tab:stats_norm} provides some example cases where the statistical tests were successful against a strict significance level of $\alpha = 0.0005$ for the p-value. Due to space limitation, we show only top-3 features per feature group.  For 52 out of 67  features engineered, the statistical analysis shows high potential of discriminative power. Nevertheless, in the rest of our experiments, and following insights from previous studies~\cite{perl2015vccfinder}, we keep all features for the learning process as some combinations may contribute to yielding an efficient classifier.

%
%
%
%

\begin{table}[!h]
\vspace{-2.0mm}
	\centering
	\caption{Statistical analysis results for top normalized features with highest discriminative potential.}
	\resizebox{1\linewidth}{!}{%
		\begin{tabular}{lccc|ccc|ccc}
			\cmidrule{2-10}
			&\multicolumn{3}{c|}{Code-fix features}& \multicolumn{3}{c|}{sec.-sensitive features}&\multicolumn{3}{c}{Text features}\\
			\cmidrule{2-10}
			& F6&F16&F24&F11&F22&F24&W2&W4&W6\\
			\midrule
			Mean for&&&&&&&&&\\
			security patches&0.120&0.038&0.110&0.004&0.006&0.350&0.360&0.360&0.350\\
			\midrule
			Mean for&&&&&&&&&\\
			other patches&0.090&0.016&0.050&0.003&0.004&0.330&0.310&0.320&0.330\\
			\midrule
			 P-value (MWW) &$5e^{-62}$&$2e^{-40}$&$4e^{-103}$&$1e^{-13}$&$1e^{-15}$&$6e^{-47}$&$2e^{-65}$&$2e^{-66}$&$7e^{-50}$\\
			\bottomrule
		\end{tabular}}	
	\label{tab:stats_norm}
\vspace{-2.0mm}
\end{table}

%

\subsubsection{Classification experiments}
The previous statistical analysis  assessed the discriminative power of engineered features with respect to security patches and the combined set of bug fixing and code enhancement patches. We propose to further assess the behaviour of one-class classification models with these features applied to the unlabeled patches.
Our experiments aim at answering two questions:
\begin{itemize}[leftmargin=*]
	\item {\em Can the features help effectively classify unlabeled patches?} We attempt to assess to what extent unlabeled patches that are flagged as security patches would constitute noise or good samples to help augment the training data of a binary classifier.
	\item {\em Are the feature categories independent and thus splittable  for a Co-Training model learning?} The choice of Co-Training as an approach is based on the hypothesis that the views are redundant. However, another constraint for the efficacy of Co-Training is that the features must be independent~\cite{nigam2000analyzing} (i.e., they do not lead to exactly the same classifications).
\end{itemize}

\begin{figure*}[!t]
\centering
\begin{subfigure}{.3\textwidth}
  \centering
  \includegraphics[width=0.6\linewidth]{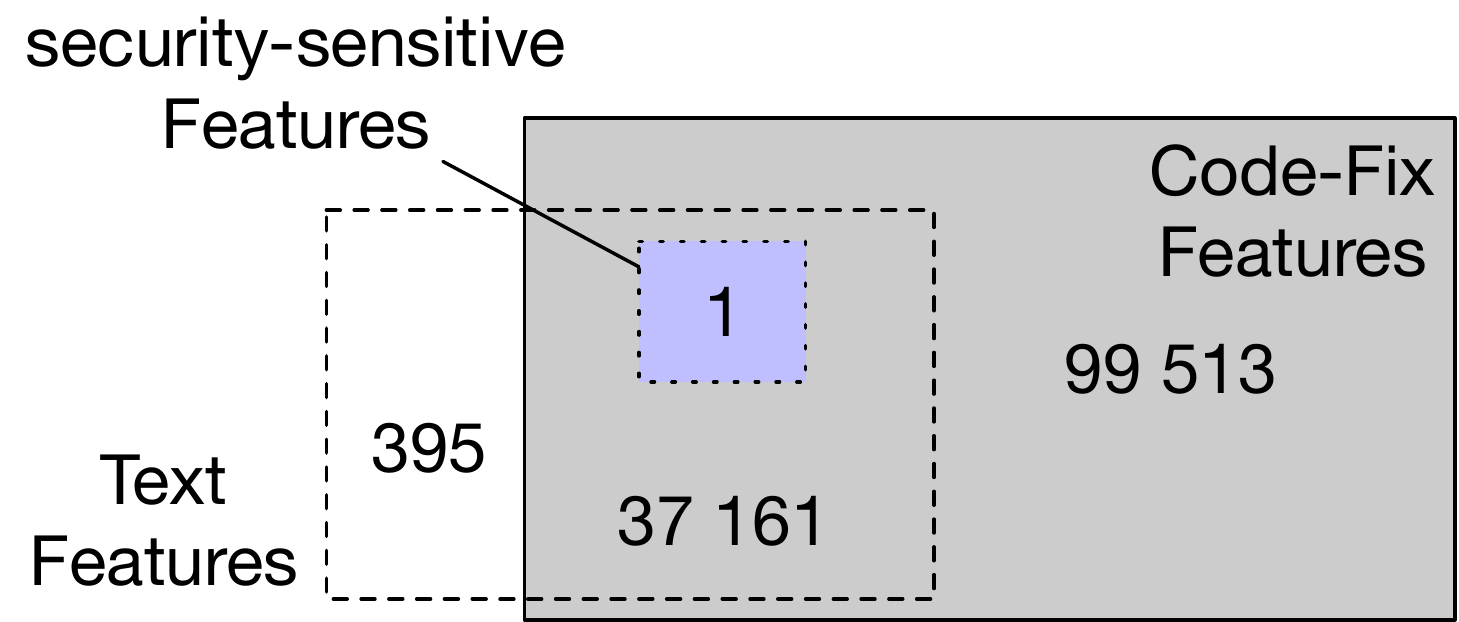}
  \caption{\normalfont Flagged Linux unlabeled patches}
  \label{fig:sub1}
\end{subfigure}%
\begin{subfigure}{.37\textwidth}
  \centering
  \includegraphics[width=0.55\linewidth]{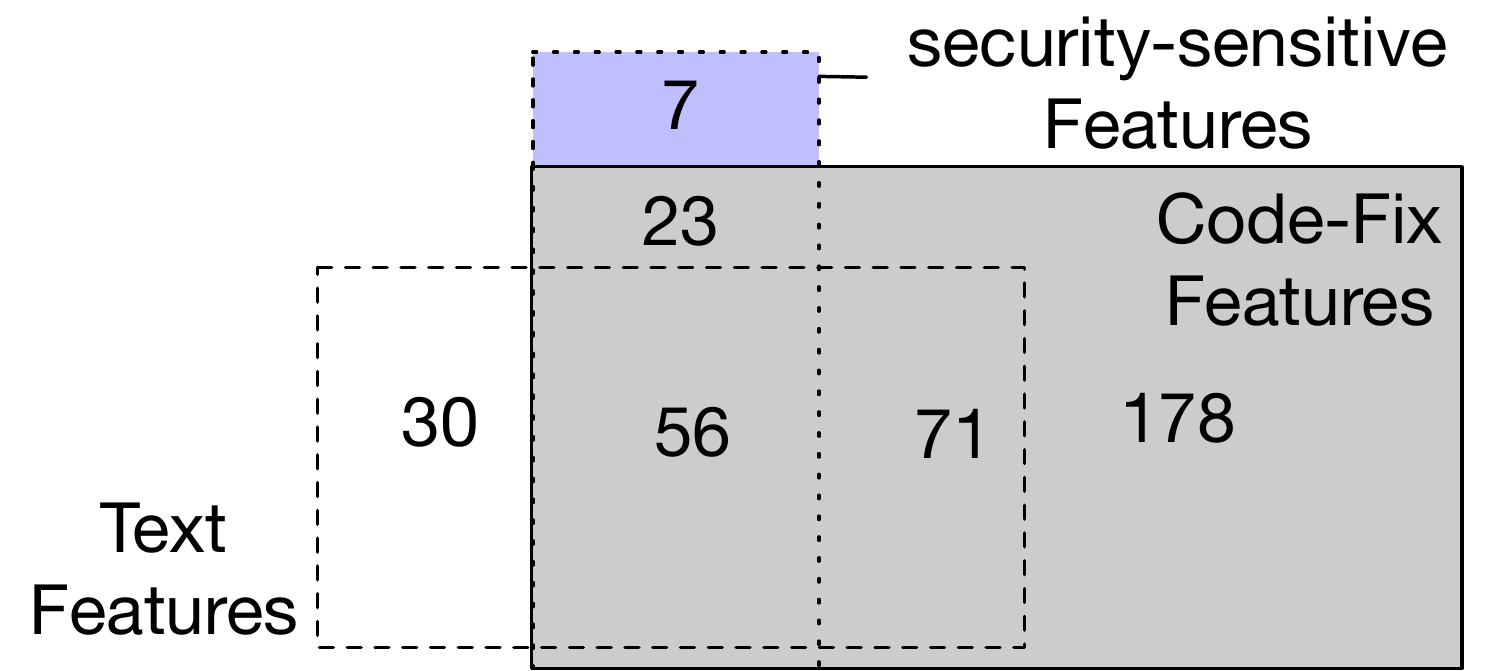}
\caption{\normalfont Flagged OpenSSL unlabeled patches}
  \label{fig:sub2}
\end{subfigure}
\begin{subfigure}{.3\textwidth}
  \centering
  \includegraphics[width=0.6\linewidth]{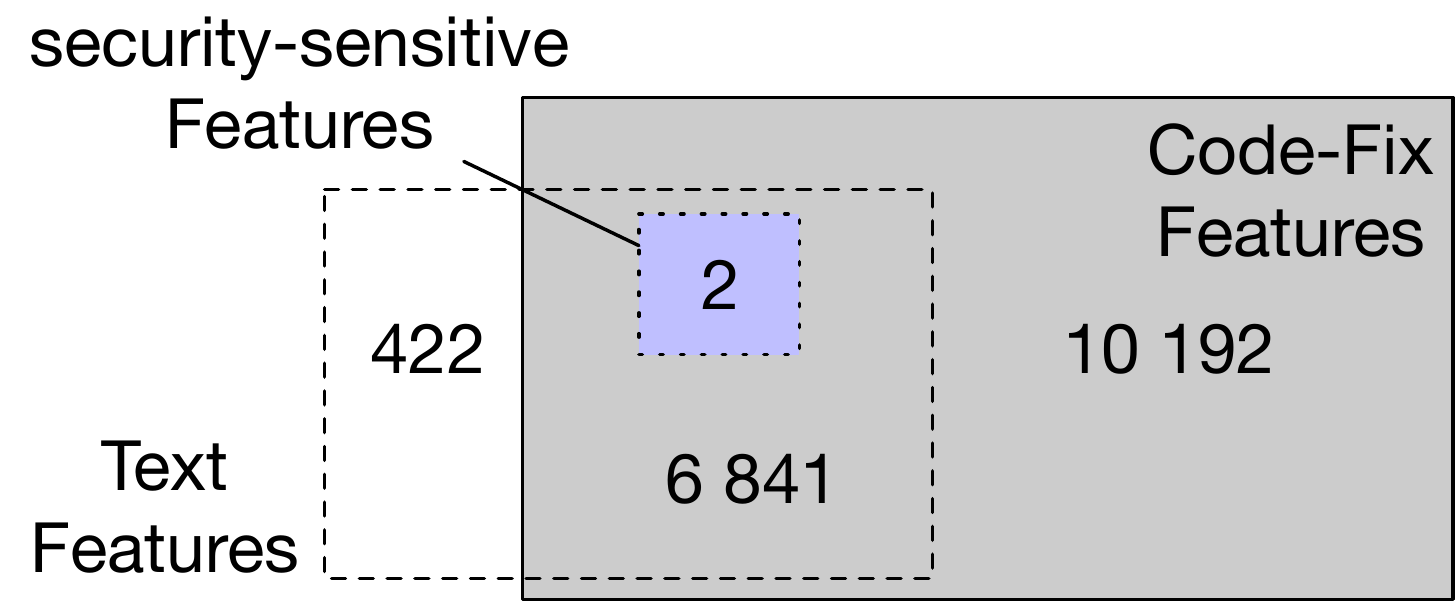}
\caption{\normalfont Flagged Wireshark unlabeled patches}
  \label{fig:oneclass-overlaps}
\end{subfigure}
\caption{Euler diagrams representing the overlaps between sets of unlabeled patches that are classified as security patches when using One-Class SVM model based on variants of feature sets.}
\label{fig:overlaps}
\vspace{-0.5cm}

\end{figure*}

\noindent
{\bf \em Features efficiency.}
Various verification problems in machine learning 
involve identifying a single class label
as a `target' class during the training process, and at prediction time make a
judgement as to whether or not an instance is a member of the target class~\cite{hempstalk2008discriminating}. In many cases, a one-class classifier is used in preference to a multi-class
classifiers, mainly because it is inappropriate or challenging to collect or use non-target data
for the given situation. In such cases, the one-class classifier is actually an {\em outlier detector} since it attempts to differentiate between data that appears normal (i.e., from the target class) 
and abnormal with respect to a training data composed only of normal data. Thus, if the features are not efficient to fully characterize the normal data in the training set, many samples classified as normal will actually be false positives and thus constitute {\em noise} in an augmented set of normal data.

Given the lack of ground truth (for unlabeled patches), we 
assess whether unlabeled patches that are flagged as security patches by a one-class classifier are noise (i.e., false positives), and thus deteriorate a binary classification performance when added to a training dataset. The comparison is done following two experiments:
\begin{itemize}[leftmargin=*,noitemsep,topsep=0pt]
	\item First, we compute accuracy, precision and recall metrics of a classical SVM \textbf{binary classifier} using the existing set of security patches as positive data and other sets of non-security (i.e., bug-fix and code enhancement) patches as negative data.
	\item Second, we augment the existing set of security patches with automatically labeled patches after applying a \textbf{one-class classifier} to the dataset of unlabeled patches. Then we use this augmented set as the positive data and redo the first experiment. This workflow is detailed in Figure~\ref{fig:workflow}.
\end{itemize}

If the features are not efficient in characterizing security patches, the one-class classifier will yield false positives and false negatives. Thus, when adding false positives to the ground truth positive data, we will be introducing noise which will lead to performance degradation. However, if the features are efficient, we will be increasing the training set and potentially leading to a better classification performance.

Equations~(\ref{eq1}) and (\ref{eq2}) provide the standard formulas for computing performance metrics, where $TP$ is the number of True Positives, $TN$ that of True Negatives, $FP$ that of False Positives and $FN$ that of False Negatives.

\vspace{-3.0mm}
\begin{equation}
\scriptsize
	Precision = \frac{TP}{TP+FP}\; ; \; Accuracy=\frac{TP+TN}{TP+TN+FP+FN}
	\label{eq1}
\end{equation}
\vspace{-4.0mm}
\begin{equation}
\scriptsize
	Recall=\frac{TP}{TP+FN}\; ; \;F1=2*\frac{Precision*Recall}{Precision+Recall}
	\label{eq2}
\end{equation}

Our experiments are performed with 10-Fold cross validation and performance is measured for the target class of security patches and only on the initial ground truth samples. Using only the initial set of security patches in the training dataset, we record an average Accuracy of 58\% (Recall = 56\%, Precision= 71\%). However, when we augment the training set with flagged unlabeled patches, we observe a clear improvement of the accuracy to 79\% (Recall = 76\%, Precision= 85\%). 

\find{The engineered features are effective for characterizing security patches. 
They can be used to collect patches for artificially augmenting a training dataset.
}


\vspace{-1.0mm}
\noindent
{\bf \em Features independence.}
\label{par:independence}
The two most closely related work in the literature~\cite{zhou2017automated,sabetta2018icsme} rely on commit text or/and code changes that they treat as simple bags of words. Nevertheless, no experiments were performed to assess the contribution and complementarity of the different information parts. We explore these contributions by evaluating the overlap among the unlabeled patch subsets that are flagged when using different feature sets. Figure~\ref{fig:overlaps} illustrates these overlaps with Euler diagrams for the different projects considered in our study. 
We note that although there are overlaps, a large portion of samples are detected exclusively with each feature set (e.g., in Linux, $99,513+395= 99,908$ patches out of $99,513+395+1+37,161= 137,070$ patches --73\%-- are exclusively detected by either code-fix features or text features). 
Nevertheless, we note that security-sensitive features are more tightly related to code-fix features (except for 7 patches in OpenSSL, all flagged patches with security-sensitive features are also flagged with code-fix features\footnote{This does not mean that security-sensitive features are useless or redundant. Patches flagged with code-fix features are scarcely flagged with security-sensitive features.}, which was to be expected given that security-sensitive features are also about ``fixing'' code).
We then conclude that {\em code-fix} features can be merged with {\em security-sensitive features} to form {\bf  code features}, which constitute a feature set that is {\bf independent} from the {\bf  text features} set. As Krogel and Schefferd demonstrated, Co-Training is only beneficial if the data sets used in classification are independent~\cite{krogel2004multi}. This insight on the sets of engineered features serves as the foundation for our model learning detailed in the following paragraphs.

\find{{\bf Code features} ({\em formed by security-sensitive features + code-fix features}) and {\bf Text features} are independent. They will represent two distinct views of the data, an essential requirement for Co-Training.}

\vspace{-3.0mm}
\subsection{Co-Training Model Learning}
\label{subsec:Co-Training}
\vspace{-1.0mm}
Experimental results described above have established that
the different features engineered provide meaningful information for the identification of security patches. Nevertheless, given the large number of these features, manual construction of detection rules is difficult. We propose to apply techniques from the area of machine learning to automatically analyze the code commits and flag those that are most likely to be delivering security patches.

In the construction of our learning-based classifier, we stress on the need for practical usefulness to practitioners. Thus,
following recommendations by authors~\cite{perl2015vccfinder} proposing automatic machine-learning approaches to support security analysts,  we 	strive to build an approach towards addressing the following challenges:
\begin{itemize}[leftmargin=*,noitemsep,topsep=0pt]
	\item Generality: Our feature engineering mixes metadata information from commit logs, which may or may not be explicit, with numerical code metrics. It is thus  important that the classifier effectively leverages those heterogeneous features to infer an accurate combined detection model.
	\item Scalability: Given that most relevant software projects include thousands of commits that must be analyzed, it is necessary for the approach to be able to operate
on the large amount of available features in a reasonable
time frame.
	\item Transparency: In practice, to be helpful for analysts, a classifier must provide human-comprehensible explanations with the classification decision. 
		For example, instead of requiring an analyst to blindly trust a black-box decision based on deep features, 
		information gain\footnote{Information gain is a metric based on entropy that allows to tell how important a given attribute of the feature set is.} 
		(InfoGain) scoring values of human-engineered features can be used as hints for manual investigation.
\end{itemize}

\subsubsection{Model Learning} 
Experiments with one-class classification have already demonstrated that it is possible to build a classifier that fits with the labeled patches in the ground truth data. Unfortunately, in our case,
a major problem in building a discriminative classifier is the non-availability of labeled data: 
the set of unlabeled patches is significantly larger than the limited dataset of labeled patches that we could collect.  A classification task for identifying security patches requires examples of both security and security-irrelevant patches. In related work from the security industry~\cite{zhou2017automated}, team members having relevant skills and experience spent several months labeling closed-source data to support the model learning. 
Since their dataset was not publicly\footnote{Our requests to obtain datasets from authors of~\cite{zhou2017automated} and ~\cite{sabetta2018icsme} remained unresponded.} available,  we propose to rely on the Co-Training algorithm to solve the non-availability problem. The algorithm was proposed by Blum and Mitchell~\cite{blum1998combining}, for the problem of semi-supervised learning where there are both labeled and unlabeled examples. The goal of Co-Training is to enhance performance of learning algorithm when only a small set of labeled examples is available. The algorithm trains two classifiers separately on two sufficient and redundant views of the examples and lets the two classifiers label unlabeled examples for each other.

\begin{figure}[!h]
\vspace{-3mm}
	\centering
	\includegraphics[width=0.81\linewidth]{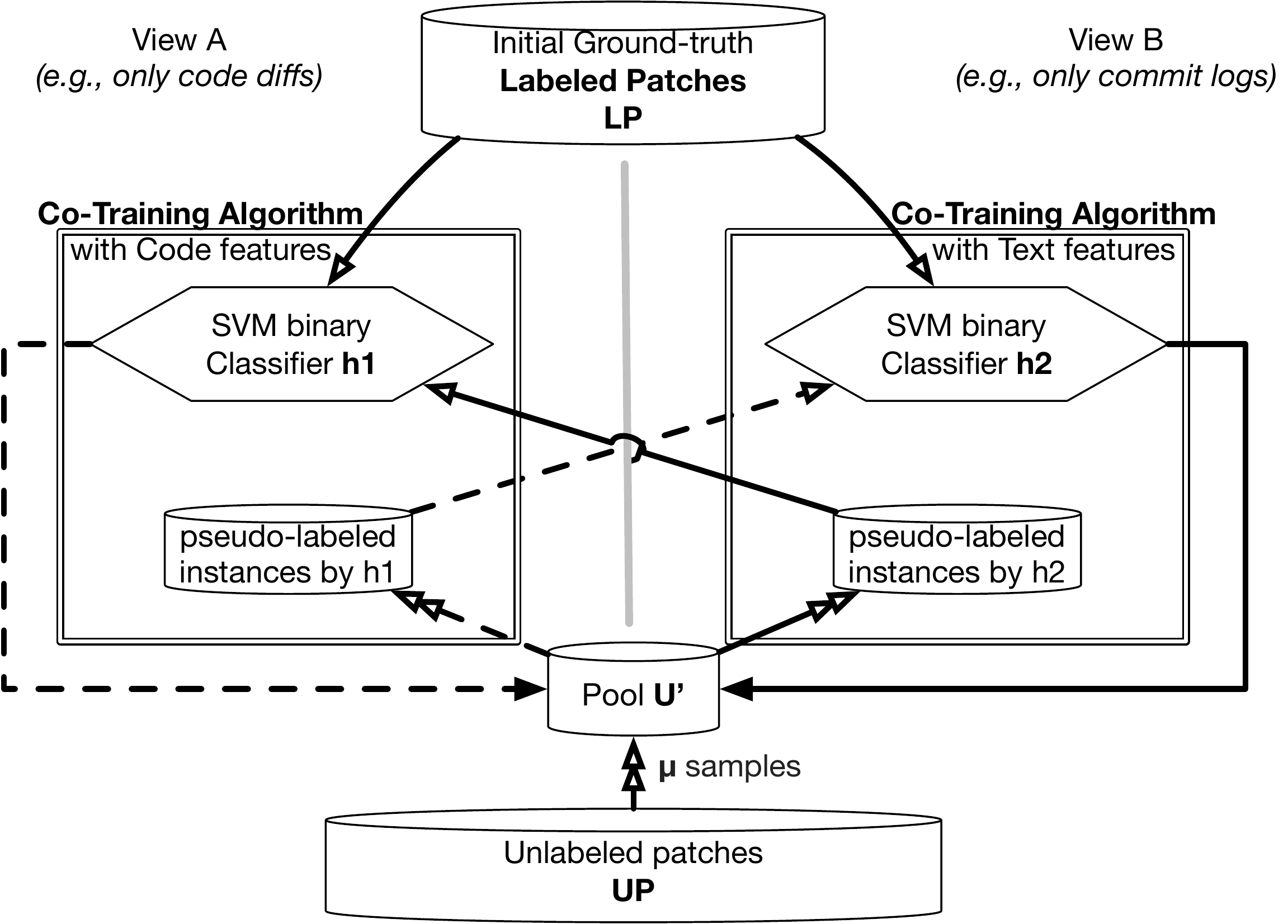}
	\caption{Co-Training learning model (cf. details in Algorithm~\ref{algo:steps})}
	\label{fig:Co-Training}
\vspace{-5mm}
\end{figure}

Figure~\ref{fig:Co-Training} illustrates the Co-Training process implemented in this work. An important assumption in Co-Training is that each view is conditionally independent given the class label. We have demonstrated in Section~\ref{par:independence} that this was the case for the different categories of features explored in this work. Indeed, Co-Training is effective if one of the classifiers correctly labels a sample that the other classifier previously misclassified. If both classifiers agree on all the unlabeled patches, i.e. they are not independent, labeling the data does not create new information.

Concretely, given a training set comprising labeled patches  and noted $LP$, and a set of unlabeled patches $UP$, the algorithm randomly selects $\mu$ samples from $UP$ to create a smaller pool $U'$, then executes the process described in Algorithm~\ref{algo:steps} during k iterations.

The overall idea behind the Co-Training algorithm steps is that the classifier $h_1$ adds examples to the labeled set which are in turn used by the classifier $h_2$ in the next iteration and vice versa. This process should make classifiers $h_1$ and $h_2$ to agree with each other after $k$ iterations. In this study we selected Support Vector Machines (SVM)~\cite{vapnik2013nature} as the internal classification algorithm for the Co-Training. SVM indeed provide tractable baseline performance for replication and comparisons against state-of-the-art work.

\subsubsection{Identification of security patches}
Eventually, when the Co-Training is stabilized (i.e., the two internal classifiers agree), the output classifier can be leveraged to classify unlabeled patches. Eventually, in this work, we consider the classifier built on the code view (which has been constantly improved due to the co-training) as the yielded classifier.

\begin{algorithm}[!h]
\scriptsize
    \SetKwInOut{Input}{input}
    \SetKwInOut{Output}{output}
    \SetKw{Return}{return}

    \Input{training set ($LP$), unlabeled data ($UP$)}
    \Input{pool $U'$ }
    \BlankLine
    \Output{$U'$: updated pool}
    \Output{$LP$: updated training set}
    \BlankLine
    \SetKwProg{Fn}{Function}{}{end}
    \BlankLine
    \Fn{getView($x$, $classifier$)}{
    	\If{$classifier$ = $C_1$} {
            	return $Text\_features(x)$
            }
            	return $Code\_features(x)$
            
	}
	    \Fn{buildClassifier($first$)}{
    	$vectors = \emptyset $\;
    	\eIf{$first$ = $True$} {
        \ForEach{$x \in LP$}{
    			$vectors = vectors \cup getView(x, C_1)$\;
            }
        }{
        \ForEach{$x \in LP$}{
    			$vectors = vectors \cup getView(x, C_2)$\;
            }
        }
        $classifier \leftarrow train\_model($SVM$, vectors)$\;
        return $classifier$\;
	}
	$h_1 \leftarrow buildClassifier(True)$;    \,\,\,\,\,\,\,\,\,$h_2 \leftarrow buildClassifier(False)$\;
	$(P_1, N_1) \leftarrow classify(h_1, U') $;  \,\,\,\,\,\,\,\,\,\,  $(P_2, N_2) \leftarrow classify(h_2, U') $\;
	$LP \leftarrow LP \cup random\_subset(\#p, P_1) \cup random\_subset(\#p, P_2)$\;
	$LP \leftarrow LP \cup random\_subset(\#n, N_1) \cup random\_subset(\#n, N_2)$\;
	$U' \leftarrow U' \cup random\_subset(\#2*(p+n), UP)$\;
	
	\caption{Steps for each Co-Training iteration. 
	}\label{algo:steps}
	
\end{algorithm}

\begin{figure*}[!t]
\centering
\begin{subfigure}{.33\textwidth}
  \centering
  \includegraphics[width=1\linewidth]{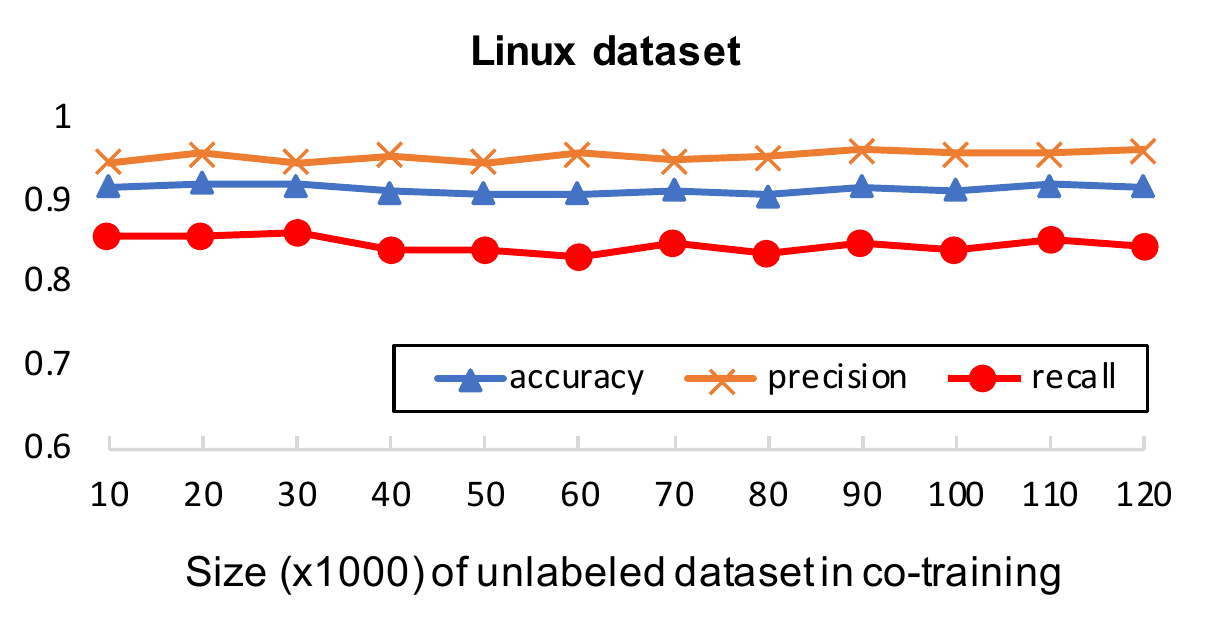}
\end{subfigure}%
\begin{subfigure}{.33\textwidth}
  \centering
  \includegraphics[width=1\linewidth]{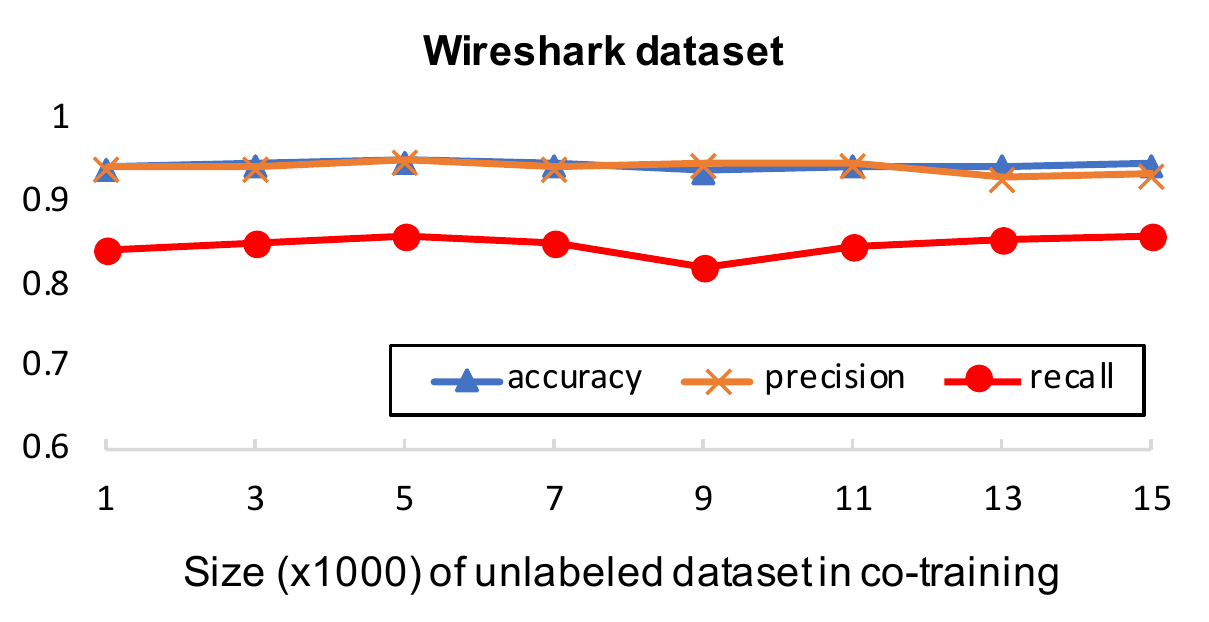}
\end{subfigure}
\begin{subfigure}{.33\textwidth}
  \centering
  \includegraphics[width=1\linewidth]{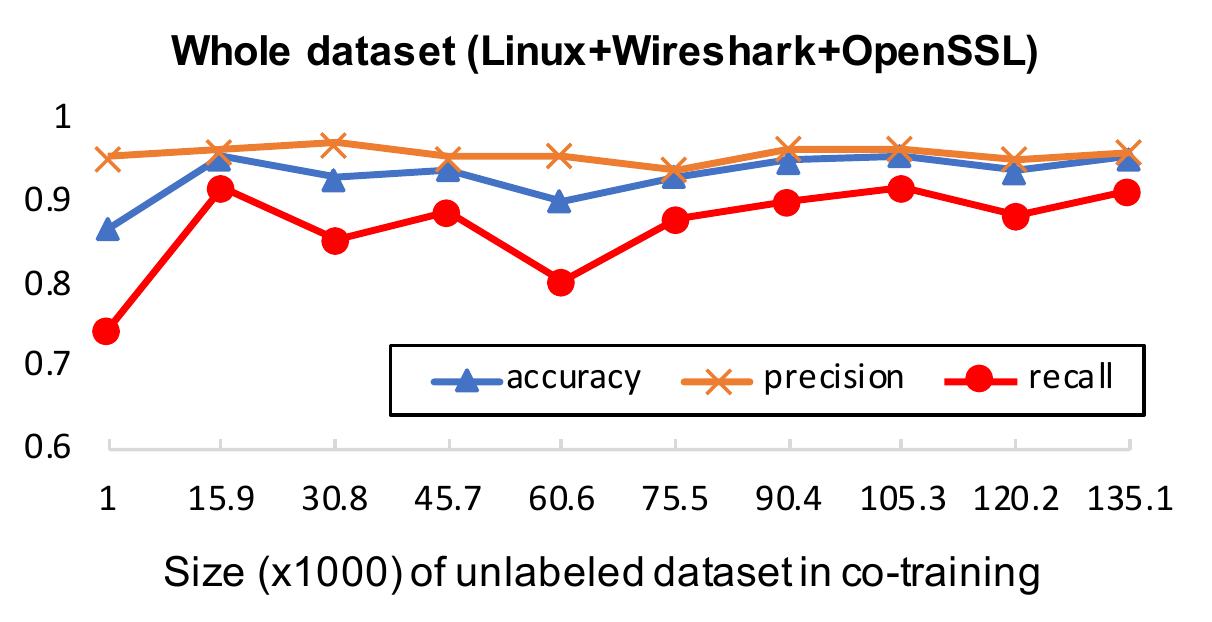}
\end{subfigure}
\caption{Precision, Recall and Accuracy metrics in benchmark evaluation with varying sizes for the unlabeled dataset.}
\label{fig:bench-results}
  \vspace{-0.5cm}
\end{figure*}


\section{Experimental Study and Results}
\label{sec:assessment}
\vspace{-1.0mm}
Our experiments aim at assessing the performance of the overall approach, detailing the impact of the Co-Training algorithm and comparing against the state-of-the-art. We investigate the following research questions:
\begin{itemize}
	\item [RQ-1.] What levels of performance can be reached by the Co-Training algorithm in the classification of patches?
	\item [RQ-2.] Can we learn to classify patches across projects?
	\item [RQ-3.] How does our Co-Training approach compare against the	state-of-the-art?
	\item [RQ-4.] Can the approach flag unlabeled patches in the wild?
\end{itemize}

\vspace{-3.0mm}
\subsection{RQ1: Classification performance}
We perform binary classification experiments to assess the performance of classifiers in discriminating between security patches (positive class) and non-security patches (negative class). We remind that, as illustrated in Figure~\ref{fig:datasets}, the non-security patches consist in the pure bug-fix patches and code-enhancement patches.  
These experiments, similarly to past studies~\cite{sabetta2018icsme,zhou2017automated,tian2012identifying}, report performance based on the ground-truth data (i.e., unlabeled patches are not considered to compute the performance score).
 
Our first experiment investigates the performance of the Co-Training approach when varying the size of the unlabeled dataset. 
In this experiment, we randomly split the labeled patch sets into two equal size subsets: one subset is used in conjunction with the unlabeled dataset for the Co-Training, while the other is used for testing. Precision, Recall and Accuracy are computed based on the test set. Figure~\ref{fig:bench-results} presents the results, showing precision measurements above 90\%, and recall measurements between 74\% and 91\%. We do not show evaluation graphs for OpenSSL dataset since this dataset included only 436 unlabeled patches. With this quantity of unlabeled data, our approach yields with OpenSSL the lowest Precision metrics at 74\%, but the highest Recall at 93\%. 
We note that, when using the whole dataset (including all projects data) the performance remains high. The best performing state-of-the-art approach in the literature for identifying security-relevant commits has reported Precision and Recall metrics at 80\% and 43\% respectively~\cite{sabetta2018icsme}. Tian et al. have also reported F1-Measure performance around 70\% for identifying bug fixing commits~\cite{tian2012identifying}, while the F1-measure performance of our approach is 89\% on average.

Our second experiment assesses the contribution of the feature set on the one hand, and of the choice of Co-Training as learning algorithm on the other hand. We replicate the SVM binary classifier proposed by Sabetta and Bezzi~\cite{sabetta2018icsme} and apply it on our labeled patches. We also build a similar classifier, however using our own feature set. 
We perform 10-fold cross validations for all classifiers and evaluate the performance of the classifier in identifying labeled security patches in the whole dataset. Results in Table~\ref{tab:algo_feature} indicate that our feature set is more effective than those used by the state-of-the-art, while the Co-Training semi-supervised model is more effective than the classical binary classification model.
 
\begin{table}[!h] 
  \vspace{-2.0mm}
\centering
	\caption{Importance$^\ast$ of Classification method and feature set}
	\resizebox{0.8\linewidth}{!}{%
	\begin{tabular}{lccc}
		\toprule
		&Precision& Recall & F1-measure  \\
		\cmidrule{2-4}
		SVM binary classification &  &  &\\
		(\em with features of Sabetta \& Bezzi ~\cite{sabetta2018icsme}) &0.60 & 0.63  &0.61\\
		\cmidrule{2-4}
		SVM binary classification& & & \\
				(\em with our feature set) & 0.85 &0.86 & 0.85\\
		\midrule
	Co-Training + SVM& & &  \\
					(\em with our feature set) &0.96 & 0.90  &0.93\\
		\bottomrule
	\end{tabular}
	}
	\label{tab:algo_feature}
	{\\\footnotesize $^\ast$Performance metrics are for classifying 'security patches'. Due to space limitation, we refer the reader to the replication package for all evaluation data.}
  \vspace{-0.3cm}
\end{table}
%

Given that our code-fix features overlap with features used by Tian et al.~\cite{tian2012identifying} for classifying bug fix patches, we present performance comparisons with the different feature sets. Results in Table~\ref{tab:resultatsTian} confirm that our extended feature set (with vulnerability-sensitive features) allows to increase performance by up to 26 percentage points. 
The performance differences between projects further confirm that the features of Tian et al.~\cite{tian2012identifying} are indeed very specific to Linux.

\begin{table}[!h] 
  \vspace{-3.0mm}
	\centering
	\caption{F-Measure Comparison: Our features vs features in~\cite{tian2012identifying}$^\ast$}
	\resizebox{0.8\linewidth}{!}{%
		\begin{tabular}{lcccc}
			\toprule
			&OpenSSL& Wireshark & Linux & Whole data \\
			\cmidrule{2-5}
			 Co-Training + SVM & & & & \\
   (\em with our feature set) &0.93&0.89 &0.94 &0.93 \\ \midrule
   		 Co-Training + SVM & & & & \\ 
   (\em with feature set of Tian \& al. ~\cite{tian2012identifying}) &0.65 & 0.71  &0.96& 0.77 \\ \midrule
   		 SVM binary classification & & & & \\
   (\em with features of Tian \& al. ~\cite{tian2012identifying}) &0.69 & 0.77  &0.99& 0.69\\ 
   
			\bottomrule
		\end{tabular}
	}
	\label{tab:resultatsTian}
	{\\ \footnotesize{This comparison serves to assess the impact of our security-sensitive features}}
	\vspace{-0.3cm}
\end{table}

\find{{\bf RQ1}$\blacktriangleright$Our approach (Co-Training + feature set) yields a highly accurate classifier for classifying patches with respect to whether they are security-relevant or not. Our performance results are above those reported by prior work for classifying patches.$\blacktriangleleft$}

 \vspace{-3.0mm}
\subsection{RQ2: Cross-project classification}
In the wild of software development projects, as reflected by the case of OpenSSL, there can be limitations in the available labeled data. Thus, it could be beneficial if practitioners can train a model by leveraging data from another project and still obtain reasonable classification performance on a distinct target project. We investigate this possibility on our datasets considering that they are written in the same programming language (C). Table~\ref{tab:cross-projects} shows the classification performance results, in terms of Recall and Precision, when training on one project and applying the model to another. We note that training on Wireshark data yields reasonable (although not optimal) performance on OpenSSL patches, while training on OpenSSL interestingly offers high performance on Linux patches. In both cases, the converse is not true. 
Variations in cross-project performances may be explained by factors such as 
coding styles differences, code base size or different security patching 
policies among projects.
Future work will investigate effects of these factors.

\begin{table}[!h]
  \vspace{-3.0mm}
\centering
	\caption{Results of cross-project classification}
	\resizebox{0.8\linewidth}{!}{%
	\begin{tabular}{llccc}
		\toprule
		&&\multicolumn{3}{c}{\em \bf Training on}\\ 
		\cmidrule{3-5}
		&& OpenSSL & Wireshark & Linux \\
		&& precision/recall & precision/recall  & precision/recall  \\
		\cmidrule{2-5}
		\parbox[t]{2mm}{\multirow{3}{*}{\rotatebox[origin=c]{90}{\bf Testing on}}} & OpenSSL& (0.93 /0.94) & 0.71 / 0.48 & 0.42 / 0.88\\
		\cmidrule{2-5}
		&Wireshark &0.53 / 0.88  & (0.93 / 0.85) &0.50 / 0.95  \\		
		\cmidrule{2-5}
		&Linux &0.89 / 0.78 & 0.45 / 0.93 & (0.95 / 0.84) \\		
		\bottomrule
	\end{tabular}
	}
	\label{tab:cross-projects}
  \vspace{-0.3cm}
\end{table}

%
%
%

\find{{\bf RQ2}$\blacktriangleright$Cross-project classification can yield comparatively good performance in some cases of combinations, such as when training on OpenSSL to classify Linux patches.$\blacktriangleleft$}


  \vspace{-3.0mm}
\subsection{RQ3: Comparison with the state-of-the-art}
While we report a F-Measure performance of around 90\%, the most recent 
state-of-the-art on security commit classification (i.e.,~\cite{sabetta2018icsme}) 
reports performance metrics around 55\%. 
Our experiments however are performed on different datasets because the 
dataset used by Sabetta \& Bezzi was not made available. 
Thus, we replicate the essential components of the best performing
approach in their work~\cite{sabetta2018icsme} (i.e., SVM bi-classification with
bag-of-words features of code and log), and can therefore compare\footnote{Note that the recorded performance of the replicated approach on our dataset is in line with the performance reported by the authors in their paper~\cite{sabetta2018icsme}.}
their approach and ours in Table~\ref{tab:comparison}.

\begin{table}[!h]
  \vspace{-3.0mm}
\centering
	\caption{Comparison of F-Measure metrics}
	\resizebox{0.8\linewidth}{!}{%
	\begin{tabular}{lcccc}
		\toprule
		&OpenSSL& Wireshark & Linux & Whole data \\
		\cmidrule{2-5}
		Our Approach & 0.93&0.89 &0.94 &0.91\textbf{}  \\
		\cmidrule{2-5}
		Sabetta \& Bezzi ~\cite{sabetta2018icsme}&0.45 & 0.45  &0.67& 0.61\\
		\bottomrule
	\end{tabular}
	}
	\label{tab:comparison}
  \vspace{-0.3cm}
\end{table}

\find{{\bf RQ3}$\blacktriangleright$Our Co-Training approach outperforms the state-of-the-art in the identification of security-relevant commits.$\blacktriangleleft$}


  \vspace{-3.0mm}
\subsection{RQ4: Flagging unlabeled patches}
Performance computation presented in previous subsections are based on cross validations where training and test data are randomly sampled. 
Such validations often suffer from the data leakage problem~\cite{ribeiro2016should}, which leads to the construction  overly optimistic models that are practically useless and cannot be used in production. 
For example, in our case, data leakage can happen if the training set includes security patches that should actually only be available in the testing set (i.e., we would be learning from the future). 
We thus propose to divide our whole dataset, with patches from all projects, following the commits timeline, and select the last year's commits as test set. 
The previous commits are all used as training set. We then train a classifier using our Co-Training approach and apply it to the 475 commits of the test set. 
To ensure confidence in our conclusions, we focus on automatically measuring the performance based only on the last year patches for which the labels are known (i.e., the patches coming from  the security patches dataset, the pure bug fix patches dataset, and the code enhancement patches dataset as illustrated in Figure~\ref{fig:datasets}).
Overall, we recorded precision and recall metrics of 0.64 and 0.67 respectively.

In a final experiment, we propose to audit 10 unlabeled patches 
flagged as security patches by a Co-Training classifier built by learning on the whole data. 
We focus on the top-10 unlabeled patches that are flagged by the classifier with the highest prediction probabilities. Two authors manually cross-examine the patches to assess the plausibility of the classification.  We further solicit the opinion of two researchers (who are not authors of this paper) to audit the flagged security patches. For each presented patch, patch auditors must indicate whether yes or no they accept it as a security patch. Auditors must further indicate in a Likert scale to what extent the associated details on the features with highest InfoGain was relevant to the reason why they would confirm the classification. Among the 10 considered patches, 5 happen to be for Linux, 3 for OpenSSL and 2 are for Wireshark.

We compute Precision@10 following the formula :
{$$Precision @k=\frac{1}{\#auditors} \sum_{i=1}^{\# auditors}\frac{\# confirmed\;patches}{k}$$}

Ideally, a security patch should be confirmed experimentally by attempting an exploit. Nevertheless, this requires extremely high expertise for our subjects (Linux, OpenSSL and Wireshark) and significant time. Instead, and to limit experimenter bias, auditors were asked to check at least whether issues fixed by the patches have similar occurrences in line with known potential vulnerabilities. For example, one of the flagged security patches is ``{fixing a memory leak}'' in OpenSSL (cf. commit {\small \tt 9ee1c83}). The literature indicates this as a known category of vulnerability which is easily exploitable~\cite{szekeres2013sok}. 

\begin{figure}[!h]
  \includegraphics[width=1\linewidth]{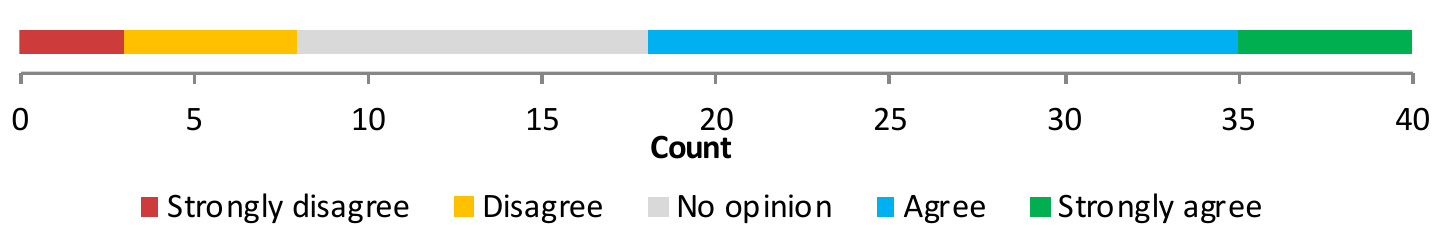}
\caption{Do the highlighted features provide relevant hints for manual review of flagged patches?}
\label{fig:likert}
\end{figure}

At the end of the auditing process, we record a Precision@10 metric of 0.55. 
Although this performance {\em in the wild} may seem limited, 
it is actually comparable to the performance recorded {\em in the lab} by the 
state-of-the-art, and is a very significant improvement over a random classifier that,
given the small proportion of security patches~\cite{ponta2019manually}, 
would almost always be wrong. 
Figure~\ref{fig:likert} indicates the distribution of the Likert scale values for the satisfaction rates indicated by the auditors for the usefulness of leveraging the features with highest InfoGain to confirm the classification. 

%

\find{{\bf RQ4}$\blacktriangleright$The approach helps to catch some silent security patches. Features with high InfoGain can be useful to guide auditors.$\blacktriangleleft$}

\section{Discussion}
\label{sec:discussion}
\noindent
{\bf \em Threats to validity.}
As with most empirical studies, our study carries
some threats to the validity.
An important threat to  {\em internal validity} in our study is 
the experimenter bias when we personally labeled 
code enhancement commits. However, we have indicated the systematic steps
for making the decisions in order to minimize the bias.
As a threat to  {\em external validity}, the generalizability of the results can be questioned since we could only manually assess a small sample set of flagged unlabeled patches. Given that our ranking is based on prediction probability, assessment of top results is highly indicative of the approach performance.
Finally, threats to {\em construct validity} concern our evaluation criteria. Nevertheless, we used standard metrics such as Precision, Recall, F-Measure and Likert scale to evaluate the effectiveness of our approach.

\noindent
{\bf \em Excluded features.} During feature extraction, we have opted to ignore information related to the author of a commit or the file where the commit occurs, as such information can lead to an overfitted model. Furthermore, we expect our classifier to be useful across projects, and thus we should not include project-specific features. In contrast, although we found that some selected features have, individually, little discriminative power, we keep them for the learning as, in combinations, they may help yield efficient classifiers.

\noindent
{\bf \em Benefit of unlabeled data.} Generally, labeling is expensive and time consuming, while unlabeled data is often freely available in large scales. Our Co-Training approach successfully leverages such data and turns a weakness in our problem setting into an essential part of the solution. 
Furthermore, it should be noted that, by construction, our dataset is highly imbalanced. Although some data balancing techniques (e.g., SMOTE~\cite{chawla2002smote}) could be used, we chose to focus our experiments on validating the suitability of our feature set with the Co-Training for semi-supervised learning. Future work could investigate other optimizations.

\noindent
{\bf \em Future work.} We plan to apply this approach to security patch identification to Java projects after collecting the necessary training data (e.g., from~\cite{ponta2019manually}). Such a classifier could then help the open source community report more vulnerabilities and fixes to security advisories. Besides SVM, which was used to ensure tractable performance comparisons with the state-of-the-art, we will investigate some Boosting algorithms. Finally,  we will consider adapting other security-sensitive features (e.g., stall ratio, coupling propagation, etc. from ~\cite{chowdhury2008security}) to the cases of code differences to assess their impact on the classification performance.

\vspace{-5mm}
\section{Conclusion}
\label{sec:conclusion}
We have investigated the problem of identifying security patches, i.e., patches that address security issues in a code base. Our study explores a Co-Training approach which we demonstrate to be effective. Concretely, we proposed to consider the commit log and the code change diff as two independent views of a patch. The Co-Training algorithm then iteratively converges on a classifier which outperforms the state-of-the-art. We further show experimentally that this performance is due to the suitability of our feature set as well as the the effectiveness of the Co-Training algorithm. Finally, experiments on unlabeled patches show that our model can help uncover silent fixes of vulnerabilities.
\newline
{\bf Availability}: We provide the dataset, scripts, and results as a replication package at {\small \bf \url{http://github.com/vulnCatcher/vulnCatcher}}

\balance
\bibliographystyle{unsrt}
\bibliography{main}

\end{document}